\documentstyle[prl,amsfonts,twocolumn,aps,epsfig]{revtex}

\begin{document}

\title{
Magneto-thermodynamics   of    the     spin-$\frac{1}{2}$     Kagom\'e
antiferromagnet}

\author{
P. Sindzingre, G.  Misguich, C. Lhuillier, B. Bernu\thanks{Laboratoire
de  Physique Th\'eorique  des Liquides, Universit\'e  P.  et M. Curie,
case 121, 4  Place  Jussieu, 75252 Paris  Cedex.  UMR 7600 of   CNRS},
L. Pierre\thanks{Universit\'e  Paris-X,  92001  Nanterre  Cedex},  Ch.
Waldtmann, H.-U.  Everts\thanks{Institut   f\"ur Theoretische  Physik,
Universit\"at Hannover, D-30167 Hannover, Germany}.  }

\maketitle

\bibliographystyle{prsty}

\begin{abstract}
In this paper, we use a new hybrid method to compute the thermodynamic
behavior of the  spin-$\frac{1}{2}$ Kagom\'e antiferromagnet under the
influence  of  a  large external magnetic  field.     We find a  $T^2$
low-temperature behavior and a very  low  sensitivity of the  specific
heat to  a strong external magnetic  field.  We display clear evidence
that this low temperature  magneto-thermal effect is associated to the
existence of low-lying fluctuating singlets,  but also that the  whole
picture  ($T^2$  behavior of  $C_{v}$    and thermally activated  spin
susceptibility) implies contribution of both non magnetic and magnetic
excitations. Comparison with experiments is made.
\end{abstract}

{\bf Introduction:}
Spin systems with strong geometrical frustration 
exhibit a generic feature: the deviation  of their spin susceptibility
from  the  Curie law occurs   at  a much   lower temperature than  for
non-frustrated  ones and  is   often  accompanied  by  very   peculiar
low-temperature    properties~\cite{r94}.    The       two-dimensional
spin-$\frac{1}{2}$ kagom\'e  antiferromagnet  (KAFM) is an  example of
such
systems.

All theoretical  and  numerical   treatments point to    a  disordered
Spin-Liquid    ground-state~\cite{e89,ze95,ce92,le93},  with a   small
spin-gap  estimated  to be  of  the  order of $1/20$   of the coupling
constant.  But   contrarily  to  what  prevails  in 1-D    or  ladders
Spin-Liquids, we have shown that  the first excitations of this system
have a spin  0.  The spectrum of  these  many-particle $S_{\rm tot}=0$
eigen-states forms  a  continuum adjacent  to the  ground-state, which
extends     smoothly      up   to   energies       higher    than  the
spin-gap~\cite{lblps97,web98}.   In  this    paper  we   report   some
thermodynamic   consequences   of this  property     and focus on  the
magneto-thermal effects which are seemingly a unique signature of this
kind of Spin-Liquids.

{\bf  Numerical  method:}
The $\frac{1}{2}$-spins on the Kagom\'e lattice interact through the Hamiltonian:
\begin{equation}
{\cal H} = J \sum_{<ij>}  {\bf S}_i. {\bf S}_j - \gamma H \sum_{i} S^z_{i}
\end{equation}
where the  double sum is limited to  first neighbors.  The temperature
$T$ and the magnetic field $H$ are measured in coupling constant units
($J=1, \gamma =1$).  Up to now two kinds of  methods were available to
obtain  thermodynamic quantities: high   temperature series and  exact
diagonalizations. The    first  ~\cite{ey94} gives  essentially  exact
results down  to temperatures of the order  of $J/2$, but unhappily no
direct information on the  range of  temperature  of interest  in this
work. The second allows  the  exact computation of  the thermodynamics
properties at all values of $T$, but for a set  of sizes limited to $N
\le 21$. For such sizes and in spite of the supposed-to-be short range
spin-spin  correlations~\cite{ce92,le93},  their    might  be     some
uncertainties on the low  $T$ thermodynamic properties related to  the
discrete nature of the finite size spectra.  To deal with larger sizes
and  in order to  obtain significant information on  the  effect of an
applied  field at low  $T$ ( $T \sim  0.05..0.2$),  we have devised an
hybrid method which  takes advantage of  both previous  approaches. An
approximate density of   states  in  each  Irreducible  Representation
(I. R.) of the complete symmetry group of  the lattice and Hamiltonian
is   reconstructed via  a maximum  entropy  procedure.  It uses  exact
diagonalization data to fix the edges of the density of states and the
six first  moments of the Hamiltonian  in each  given subspace. In the
$N=36$  sample there  are  264 different I.R.   and thus the procedure
involves a  very large amount  of exact  results. Most  of the low $T$
physics comes from the relative position, weight  and general shape of
the different  I.R., which are an  essential input of  our method. The
knowledge  of  these quantities for   each I.R.  of   $SU(2)$ allows a
straightforward computation of  the effect of  any magnetic field.  We
have checked on different frustrated  spin problems that this approach
gives reliable information  on the intermediate  temperature range and
on multiple-peak structures~\cite{greg}.

{\bf Numerical results:} {\it High temperature range down to $T=0.2$:}
Numerical  results are summarized   in Fig.~1, together  with the Pade
approximant    of  the   high-temperature   series  of     Elstner and
Young~\cite{ey94}.     In  this  range    of   temperature,  the  spin
susceptibility $\chi$ shows a shoulder at $T \sim  1$ and the specific
heat $C_v$  a  high $T$ peak   around $T=2/3$.  The  Pade approximants
cannot be distinguished from the   $N=18$ results down to $T=0.4$  and
then  merge  smoothly with the exact  $N=24$  low $T$ results. We thus
infer  that   above $T=0.4$ these   different data   give  a very good
approximation of  the thermodynamic limit.  It  should be noticed that
there is only $ 50 \%$ of the total entropy in the high $T$ peak above
$T=0.2$, whereas   the  N\'eel   ordered  Heisenberg magnet    on  the
triangular lattice develops $ 50 \%$ of the total entropy above $T=1$,
and only $ \sim  6\%$ below $T=0.2$.  The low  $T$ entropy is indeed a
distinctive  mark  of exotic  spin liquids~\cite{mblw98}  and the KAFM
represents an extreme situation.

{\it From $T=0.2$ down to $ T \sim 0.2 \Delta$:} when cooling down the
KAFM, one encounters  a second  energy  scale $\Delta$ related to  the
spin-gap~\cite{note-delta}.  In  the range of sizes presently studied,
the gap  is  still  decreasing~\cite{web98} and $\Delta$   varies from
$\Delta_{18}=0.104$  to    $\Delta_{36}=0.074$     which explains  the
evolution   of $C_v$  in   Fig.~2, 3.   From the   study  of spin-spin
correlations~\cite{ce92,le93} and  of   finite  size effects   on  the
spin-gap one might expect that the $N=36$ sample  is not very far from
the  thermodynamic limit.    Remembering that  $ 50  \%$  of the total
entropy has to be accounted for below $T=0.2$, one expects the low $T$
peak in $C_v$ to remain for $N\rightarrow\infty$.

The  spin-gap clearly    explains   the   thermally  activated    spin
susceptibility  $\chi$.  {\it However,   $C_v$ (in contrast to $\chi$)
does  not decrease exponentially in the  gap.}  For the larger samples
($N=18-36$), $C_v$ below  the peak is well  fitted by a $T^{2}$ law in
the range $ T=0.3  -0.6\Delta$ (Fig.~2).  The non exponential behavior
of $C_{v}$ is indeed due to the presence of  singlets in the gap, but,
as can  be   seen  in  Fig.~3, higher   spin channels  do   contribute
approximately to one half of $C_{v}$ in this  range of temperature and
the $T^{2}$ behavior appears as a cooperative  effect due to different
spin channels.  This fact contradicts  an intuitive expectation:  from
our  knowledge of fully gapped   systems one would  naively expect  an
absence of contribution of  the  triplets for temperatures  lower than
the   spin-gap. In fact   $C_{v}$,   at temperature  $T$,  essentially
measures the energy fluctuations  around the average energy $e(T)$. In
a  fully gapped  system,   when $T$ is  lower  than  the spin-gap, the
ground-state alone  is   populated  and the  energy   fluctuations are
thermally activated.  In the present case, the thermal population at a
temperature  $T$ is dominated  by the  singlet  states with  an energy
$\sim e(T)$: this large number of relay levels in the spin-gap and the
very large number of low   lying triplets are  at  the origin of  this
surprising behavior.

There has been in  the past many speculations  about the origin of the
$T^2$ behavior in $SrCrGaO$ and  in the jarosites  : simple or nematic
spin-waves  modes  have  been   suggested~\cite{cc91,rcc93}.  Strictly
speaking the presence of  a spin-gap precludes  such an explanation in
the spin-$\frac{1}{2}$ case, but  the small spin-gap  is a  proof that
the  spin-$\frac{1}{2}$  KAFM is  relatively  near  a quantum critical
point.  An alternative explanation  can also be featured  by focussing
on     the     Resonating-Valence-Bound         picture       of   the
singlets~\cite{rk88,ze95,m98}:   in  such  a    picture the modes  are
associated to the dimer-dimer long range orientational order (resonons
of  Rokhsar and Kivelson~\cite{rk88}).   Such a  picture gives a $T^2$
behavior  of $C_v$ provided  that  the dimer-dimer structure factor is
linear in k  at  small wave-vectors.  This  would  imply  an algebraic
decrease of the   dimer-dimer correlations.   Regarding the  range  of
sizes   available,  this  hypothesis cannot    be completely discarded
(dimer-dimer correlations at distance 6 are of  the order of $10^{-2}$
in the  spin-$\frac{1}{2}$ KAFM~\cite{le93},   whereas they  are about
$10^{-4}$ in the M.S.E.~\cite{mblw98}).   In  this last hypothesis,  a
global picture involving  both singlet and magnetic excitations should
be worked out.

{\it Temperatures   lower  than $0.2  \Delta$:}   The $C_v$ curves for
$N=18$ and  $N=36$ are  numerically exact  in this  temperature range.
The entropy of the singlet excitations  in the spin-gap ($\sim 15 \%$)
is certainly a   significant information in   the thermodynamic limit.
But the  low-temperature peak (dots in Fig.~1,  2 and 3) is unphysical
and   a consequence of  the discreteness  of  the spectra.  Large size
effects  in  the lowest  part  of  the energy  spectrum   preclude its
extrapolation  for $N\rightarrow\infty$,  in particular, they  make it
impossible to decide the  exact value of the  entropy at $T=0$ and the
very low $T$ behavior of $C_v$ in the thermodynamic limit.

{\bf Comparison  with  experimental  results:}   The  comparison  with
experimental  results   is     only   indicative, insofar   as    only
pseudo-spin-1~\cite{wkyoya97},
spin-$\frac{3}{2}$~\cite{olitrp88,rec90,ukkll94,lbar96}            and
spin-$\frac{5}{2}$~\cite{kklllwutdg96,whmmt98} Kagom\'e lattices  have
been realized and studied in the laboratory.

{\it Pseudo  Spin-1:} The organic   compound $m-MPYNN.BF_4$ studied by
Wada  and    coworkers~\cite{wkyoya97} may    behave  at    low enough
temperatures  as a spin-$1$ KAFM. $\chi$  has been measured in a large
range  of   temperature. There   are similarities  with    the present
spin-$\frac{1}{2}$ results:  a shoulder  at $T  \sim  1$, a  thermally
activated behavior with  a spin-gap of  the order of $7.7\;  10^{-2}$.
It might be noticed that the experimental gap is large enough compared
to the spin-$\frac{1}{2}$ case; half  integer and integer spin systems
might indeed behave differently.

{\it Spin-$3/2$ systems:}
We will concentrate  here on $SrCrGaO$ which
displays   the  most   convincing    example of   a   non-conventional
low-temperature    behavior~\cite{rec90,ukkll94,lbar96}.     From  the
magnetic point of  view, it is  not clear whether  $SrCrGaO$ is a good
representative of a system of Kagom\'e planes (with intervening dilute
triangular Heisenberg planes)  or whether it is  better described as a
stack  of   2-dimensional  slabs   of  pyrochlore.   However,   on the
qualitative level these  two theoretical models  display similarities,
absence of long  range order, resonating valence-bond ground-state, so
that a  comparison of the experimental  results for $SrCrGaO$ with our
numerical results might be justified~\cite{cl98}.

The double peak structure in $C_v$ has not be seen in these compounds:
this is not a  surprise at the  light of present results. In $SrCrGaO$
the Curie-Weiss temperature  is of the  order of $500K$,  the coupling
constant of equation (1) is   thus of the  order  of $100K$ and   from
results shown in Fig.~3 we  expect the low-$T$  peak  to appear at  $T
\sim 0.05J \sim  5K$ or below.  On the  other  hand the scale  for the
semi-classical behavior  is given by $S(S+1)J  $: we thus predict that
the high $T$ peak   in  $SrCrGaO$ should   appear around $350K$,   but
experimental data  only  extend to  $40K$.  We  might  notice  that in
$SrCrGaO$  as in the $S=\frac{1}{2}$ numerical  data, the low-$T$ peak
in   $C_v$   accounts for half   of  the   total  entropy of  the spin
system~\cite{r99}.  Endly the  $T^2$  behavior of $C_v$  is consistent
with the experimental data.

The  stronger  argument  in   favor   of  a parentage   between    the
spin-$\frac{1}{2}$ KAFM  and $SrCrGaO$ is  {\it  the quasi absence  of
sensitivity of $C_v$ to applied magnetic fields  $H$ as large as twice
the temperature} reported by Ramirez {\it et al.}~\cite{r99}.

This  is a highly surprising result,  which cannot be explained in any
of the following  hypothesis: N\'eel long  range  order, ordinary spin
glass, Spin-Liquid with a full gap both  for magnetic and non magnetic
excitations.

i) Since a  N\'eel-ordered system spontaneously breaks  the rotational
invariance,   the  ground-state   (and  the  first   excitations)  are
superpositions  of  quasi-degenerate eigenstates  with  {\em different
total  spin}.  In  such  a case,  $\chi$  goes to  a   constant at low
$T$. Though the  magnetization per spin  vanishes, the typical spin is
not  zero: $S_{\rm  tot}^2\sim  N\chi(T=0)$.  This  gives  a  {\em $H$
dependent} $C_v$.

ii) On the  other hand in  a fully gapped Spin-Liquid,  as soon as the
applied field is  large enough to  match the gap, $C_v$  is completely
washed out, as for example in the Spin-Peierls compounds.

In   $SrCrGaO$ a field   of 8 to  11  Tesla is  large enough to excite
magnetic excitations (as can be seen in looking at the non zero values
of $\chi$  at equivalent  temperature) and  nevertheless  there is  no
effect on $C_v$. We show on the spin-$\frac{1}{2}$ example below, that
the presence of an important background of non-magnetic excitations is
enough to drastically decrease the effect of the field on $C_v$.

The magneto-thermal effect  on spin-$\frac{1}{2}$ KAFM is displayed in
Fig.~3   for a  field  equal to   the  temperature  of the  peak $T_p$
(corresponding approximately to  the  6 Tesla experiments   of Ramirez
{\it et al.}~\cite{r99}).

i)  The contribution of singlet  excitations to $C_v$  is displayed on
the same graph. In the interesting  range of temperature, the singlets
only account for half of $C_v$.

ii) The higher  spin channels are responsible  for the missing half of
$C_v$. These  higher  spin  channels  are indeed  fully  sensitive  to
$H$. If   these   magnetic levels  were   separated  from the  singlet
ground-sate by a spin-gap void of  singlet excitations the decrease of
the $C_v$ peak   would be 3 times  larger  than it is   in the present
situation.    Here again, the presence  of  the background of singlets
brings a great  change in the  fluctuation spectrum of this system and
explains its paradoxical low sensitivity to $H$.

iii) The effect of a field  of the order  of $T_p$ seems larger in the
spin-$\frac{1}{2}$ KAFM than in $SrCrGaO$:  this suggests that in this
range  of  $T$  the  relative weight  of   the  singlets is  larger in
$SrCrGaO$ than it is for spin-$\frac{1}{2}$.  We suggest the following
conjecture: $SrCrGaO$, as  a  spin-$\frac{3}{2}$ system, is  closer to
the quantum critical transition than the spin-$\frac{1}{2}$ KAFM.  Its
spin-gap,  if  not zero  is    certainly  smaller than  that of    the
spin-$\frac{1}{2}$ KAFM, and there may  be a collapse of the different
low-energy    scales appearing in   the  low $S_{tot}$  sectors of the
spin-$\frac{1}{2}$ system  (see Fig.~3)   .  In  such  a  situation we
expect a weaker sensitivity of $C_v$ to $H$.

An important issue in experimental systems is the problem of dilution.
We  have checked the      effect    of quenched  disorder   on     the
spin-$\frac{1}{2}$ KAFM  by including 1 or 3  non magnetic  sites in a
$N=21$    sample. The  general qualitative    properties  of the  KAFM
(thermally activated $\chi$, low-$T$ peak in $C_v$ and singlets in the
spin-gap) do  not  disappear   with such  dilutions! This   seems   an
interesting indication  that  the thermodynamic features  we have been
describing in this paper are rather robust.

{\bf Conclusion:} In this paper we describe the results of a numerical
investigation  of      the  magneto-thermal    properties    of    the
spin-$\frac{1}{2}$  KAFM.  We show that the   specificities of its low
energy spectrum induce a very low sensitivity  of the specific heat to
the external magnetic field. Such an effect is a very strong signature
of  a large weight  of  singlet excitations  and  would not be present
neither in a N\'eel ordered system, in  a standard Spin-Glass nor in a
dimerized Spin-Peierls like system.  We have verified that this effect
is qualitatively  insensitive  to quenched  disorder.  The qualitative
similarities to Ramirez {\it et al}  results on $SrCrGaO$ are probably
more than a coincidence but a new independent proof advocating for the
presence of low lying  fluctuating singlets in this spin-$\frac{3}{2}$
compound. The precise nature  of the  lattice (Kagom\'e or  pyrochlore
slab)  might  not  be a central  issue.  We  have preliminary  results
showing that similar  Spin-Liquid behavior does  appear  on many other
lattices. We now suspect this behavior to be  a generic feature of one
class of Spin-Liquids.

Acknowledgments: We thanks A. Ramirez for discussion and communication
of his results a long time before publication.  We have benefited from
very   interesting   discussions with    J.P.  Boucher,   N.  Elstner,
M. Gingras, S. Kivelson  and P. Mendels.  Computations  were performed
on C98 and  T3E at the Institut de  D\'eveloppement  des Recherches en
Informatique  Scientifique  of C.N.R.S.  under contracts 984091-980076
and on the T3E at NIC J\"ulich.

\begin{figure}
\centerline{\psfig{figure=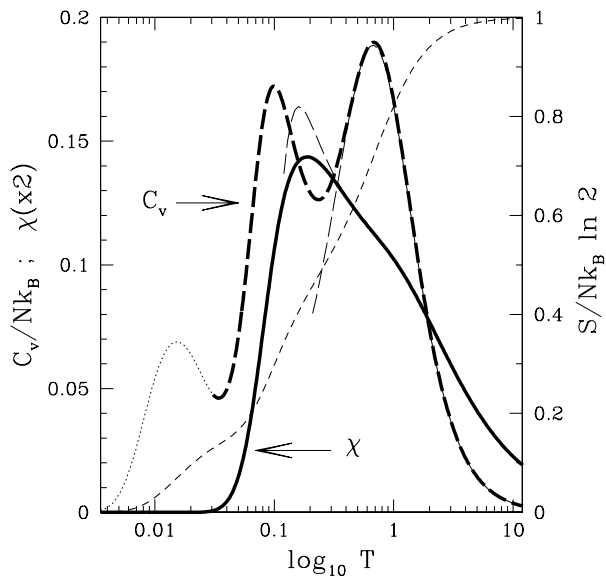,width=8.0cm,angle=0}}
    \caption[99]{Specific heat $C_{v}$, spin susceptibility $\chi$ and
    entropy  $S$   of   the  $N=18$ Kagom\'e   antiferromagnet  versus
    temperature $T$.   Dashed  line:  $C_{v}/Nk_{B}$; continuous line:
    $\chi\times  2$; short-dashed    line: $S/k_{B} \ln2$.  The  light
    dashes represent  the Pade approximants  to $C_{v}$  and $\chi$ of
    Ref.~\cite{ey94}: they cannot be distinguished from the $N=18$ and
    $N=24$ results  down to $T=0.4$.  For $T \alt 0.03$  the numerical
    results are exact but  cannot be informative on  the thermodynamic
    limit, because of the finite size effect. } \label{Fig1}
\end{figure}
\begin{figure}
\centerline{\psfig{figure=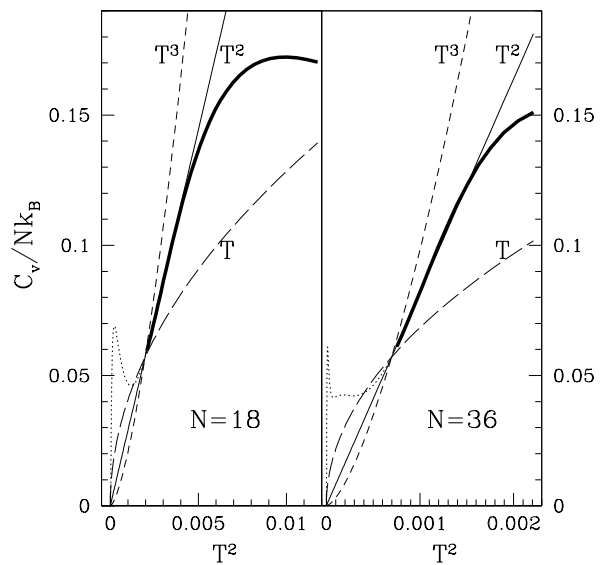,width=8.0cm,angle=0}}
    \caption[99]{ Low-temperature specific  heat $C_v$ (full lines for
    $T\gtrsim  0.3\Delta$, dotted lines  below ) versus  the square of
    the  temperature ($N=18$ and $36$ samples).   $C_v$ is well fitted
    by a $T^2$ law  (light line) in the  range $ T=0.3 -0.6\Delta$ and
    deviates from higher or lower power laws.  } \label{ Fig2}
\end{figure}

\begin{figure}
\centerline{\psfig{figure=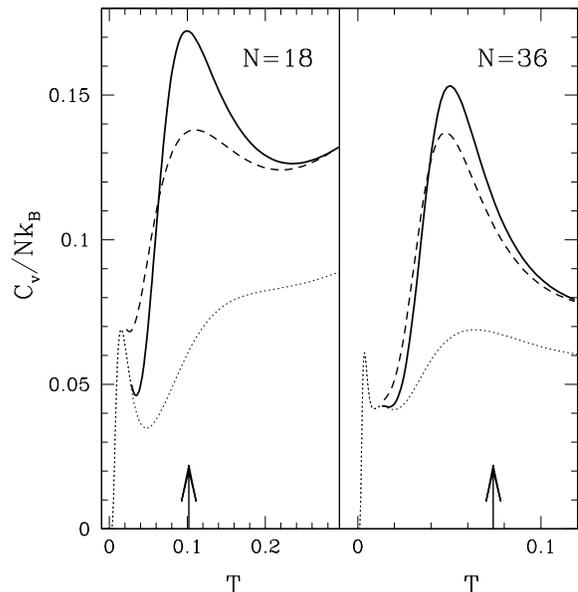,width=8.0cm,angle=0}}
    \caption[99]{ Contribution of the singlet ($S_{\rm tot}=0$) states
    (dotted lines) to the total specific heat  (full lines) and effect
    of a  magnetic field $H$ equivalent to  the 6  Tesla experiment of
    Ramirez {\it   et  al.}~\cite{r99}  (dashed  lines).   The  arrows
    indicate the value of $\Delta$ for each sample.  } \label{ Fig3}
\end{figure}

\begin{thebibliography}{10}

\bibitem{r94}
A.~P. Ramirez, Ann. Rev. Mater. Sci {\bf 24},  453  (1994), P. Schiffer and
A.~P. Ramirez, Comments Condens. Matter Phys. {\bf 18}, 21 (1996) and references
  therein. 

\bibitem{e89}
V. Elser, Phys. Rev. Lett. {\bf 62},  2405  (1989).

\bibitem{ze95}
C. Zeng and V. Elser, Phys. Rev. B {\bf 51},  8318  (1995).

\bibitem{ce92}
J. Chalker and J. Eastmond, Phys. Rev. B {\bf 46},  14201  (1992).

\bibitem{le93}
P. Leung and V. Elser, Phys. Rev. B {\bf 47},  5459  (1993).

\bibitem{lblps97}
P. Lecheminant {\it et~al.}, Phys. Rev. B {\bf 56},  2521  (1997).

\bibitem{web98}
C. Waldtmann {\it et~al.}, Eur. Phys. J. B {\bf 2},  501  (1998).

\bibitem{ey94}
N. Elstner and A.~P. Young, Phys. Rev. B {\bf 50},  6871  (1994).

\bibitem{greg}
G.~Misguich, Ph.D. thesis,  Paris VI,
France (1999).

\bibitem{mblw98}
G.   Misguich,    B.   Bernu,  C.    Lhuillier,   and   C.  Waldtmann,
Phys. Rev. Lett. {\bf 81}, 1098 (1998) and Phys. Rev. B {\bf 60}, 1064
(1999).

\bibitem{cc91}
P. Chandra and P. Coleman, Phys. Rev. Lett. {\bf 66},  100  (1991).

\bibitem{rcc93}
I. Ritchey, P. Chandra, and P. Coleman, Phys. Rev. B {\bf 47},  15342  (1993).

\bibitem{m98}
F. Mila, Phys. Rev. Lett. {\bf 81},  2356  (1998).

\bibitem{rk88}
D. Rokhsar and S. Kivelson, Phys. Rev. Lett. {\bf 61},  2376  (1988).

\bibitem{wkyoya97}
N. Wada {\it et~al.}, J. Phys. Soc. Jpn. {\bf 66},  961  (1997).

\bibitem{olitrp88}
X. Obradors {\it et~al.}, Solid State Commun. {\bf 65},  189  (1988).

\bibitem{rec90}
A. Ramirez {\it et~al.},   Phys. Rev. Lett. {\bf 64},  2070
  (1992).

\bibitem{ukkll94}
Y. Uemura {\it et~al.}, Phys. Rev. Lett. {\bf 73},  3306  (1994).

\bibitem{lbar96}
S.-H. Lee {\it et~al.}, Europhys. Lett {\bf 35},  127  (1996).

\bibitem{kklllwutdg96}
A. Keren {\it et~al.}, Phys. Rev. B {\bf 53},  6451  (1996).

\bibitem{whmmt98}
A.~S. Wills {\it et~al.}, Europhys. Lett {\bf 42},  325  (1998).

\bibitem{cl98}
B. Canals and C. Lacroix, Phys. Rev. Lett. {\bf 80},  2933  (1998).

\bibitem{r99}
A.~P. Ramirez, B. Hessen and M. Winklemann (to appear in Phys. Rev. Lett.)

\bibitem{note-delta}
For each   size,  this energy   scale  $\Delta$ is  obtained  from   a
thermodynamic  approach  of the energy  per  spin $e$ vs magnetization
$m=S_{tot}/N$: $e(m)\approx e(0)+\Delta m+ \alpha m^{2}$ at small $m$.
\end{thebibliography}
\end{document}